\documentclass[a4paper,12pt]{article}
\usepackage{graphicx,rotating,hyperref,slashed,amsmath,xcolor,amssymb,amsfonts,colortbl,cite,subcaption}
\usepackage[totalheight = 23cm, totalwidth = 17cm]{geometry}
\usepackage{physics,multirow}

\hypersetup{colorlinks, bookmarksopen, bookmarksnumbered,
citecolor=teal, linkcolor=orange, pdfstartview=FitH, urlcolor=cyan}

\newcommand{\mpl}{m_{\rm Pl}}

\newcommand{\calO}{\mathcal O}
\newcommand{\calR}{\mathcal R}

\begin{document}

\begin{titlepage}

\rightline{\footnotesize{APCTP-Pre2026-016}}

\begin{center}

\vskip 3em

{\LARGE \bf 
Algebraic classification of 
\vspace{0.3em}\\ 
single-field inflationary dynamics
}

\vskip 3em

{\large
Jinn-Ouk Gong$^{a,b}$
and
Wei-Chen Lin$^{a}$
}

\vskip 0.5cm

{\it
$^{a}$Department of Science Education,  Ewha Womans University, Seoul 03760, Korea
\\
$^{b}$Asia Pacific Center for Theoretical Physics, Pohang 37673, Korea
}

\end{center}

\vskip 1.2cm

\begin{abstract}

We develop an algebraic classification of canonical single-field inflationary dynamics using an exact depressed cubic equation for the inflaton velocity. Its root structure identifies a branch-swap boundary distinct from the limiting boundary of downhill motion, where ultra-slow-roll inflation naturally emerges. Independently, the conservation of momentum conjugate to the comoving curvature perturbation in the long-wavelength limit provides a criterion for the growth or decay of the gauge-invariant difference between comoving and uniform-density curvature perturbations. Combining these criteria, we construct a unified dynamical chart in terms of the Hubble and potential slow-roll parameters, illustrating its applicability to constant-roll inflation and simple toy models. Our classification shifts the emphasis from assigning models to fixed categories toward understanding the dynamical regimes traversed by inflationary trajectories and their implications for primordial perturbations.

\end{abstract}

\end{titlepage}

\newpage

\section{Introduction}
\label{sec:intro}

Inflation~\cite{Guth:1980zm,Linde:1981mu,Albrecht:1982wi} provides a compelling framework for explaining the initial conditions of the hot big bang universe, as supported by observations of the cosmic microwave background~\cite{Planck:2018vyg}. It also provides a natural mechanism for generating primordial cosmological perturbations from quantum fluctuations~\cite{Mukhanov:1981xt,Guth:1982ec,Hawking:1982cz,Starobinsky:1982ee,Bardeen:1983qw}. Despite its conceptual simplicity, a wide variety of inflationary models have been proposed~\cite{Martin:2013tda}, motivating attempts to organize them according to their underlying physical properties and observational predictions. A familiar example is the classification into large-field, small-field, and hybrid models, often discussed in terms of the scalar spectral index and tensor-to-scalar ratio under the usual slow-roll (SR) assumptions~\cite{Dodelson:1997hr}. The inflationary flow approach provides another systematic framework for investigating model predictions and their fixed-point structure~\cite{Kinney:2002qn}. Such classifications are useful for connecting theoretical models to observations. However, they do not directly characterize the full range of inflaton dynamics beyond SR or distinguish the background evolution from the associated large-scale behaviour of curvature perturbation.

A notable example is ultra-slow-roll (USR) inflation~\cite{Tsamis:2003px,Kinney:2005vj}, where an approximately flat potential leads to a rapid decrease of the inflaton velocity, $\dot\phi\propto a^{-3}$, and hence the SR parameter $\epsilon$ decays as $\epsilon\propto a^{-6}$ [see \eqref{eq:1st-SR}]. Moreover, unlike ordinary SR inflation, the would-be decaying mode of the comoving curvature perturbation $\calR$ can grow on super-horizon scales, potentially producing a substantial enhancement of the primordial power spectrum~\cite{Leach:2001zf,Byrnes:2018txb}. Such behaviour is not restricted to the exactly flat-potential limit, and departures from conventional SR dynamics can occur across a broader range of inflationary regimes. Moreover, the comoving curvature perturbation $\calR$ and the curvature perturbation on uniform-density hypersurfaces $\zeta$, which coincide in the usual large-scale SR attractor limit, need not remain equivalent during non-attractor evolution. Their difference offers a gauge-invariant characterization of an aspect of the large-scale perturbation dynamics. This raises a natural question: Can the background inflaton dynamics and the large-scale behaviour of curvature perturbations be classified systematically, without assuming SR or specifying a particular potential? Such a classification would be especially useful for understanding transitions between different inflationary phases and identifying the dynamical origin of super-horizon behaviour.

Previously, several approaches have explored complementary aspects of inflationary dynamics. Conventional inflationary zoology and flow methods organize potential properties, SR hierarchies, and observational predictions~\cite{Dodelson:1997hr,Kinney:2002qn}. Beyond SR, constant-roll inflation identifies solutions with a fixed inflaton acceleration~\cite{Motohashi:2014ppa}, while the phase-space stability of SR and USR backgrounds has been investigated for arbitrary potentials~\cite{Pattison:2018bct}. More recently, distinct Hamilton-Jacobi solution branches have been employed to describe long-wavelength perturbations across SR-USR transitions~\cite{Prokopec:2025uvz}. An earlier, mathematically different approach used catastrophe theory in a Hubble-parameter representation to classify inflationary potentials and critical points~\cite{Kusmartsev:1994pz}. Exact relations between Hubble-flow and potential SR parameters are also known (see, e.g.~\cite{Planck:2015sxf}), as are relations between comoving and uniform-density curvature perturbations~\cite{Romano:2015vxz,Vallejo-Pena:2019lfo}. These studies motivate a complementary approach that combines the algebraic structure of the exact background equations with a gauge-invariant characterization of large-scale perturbation evolution.

In this article, we identify such a structure in the form of an exact depressed cubic equation for the inflaton velocity. Its root structure distinguishes two positive algebraic velocity branches, whose ordering changes when the discriminant of the depressed cubic vanishes. Importantly, this branch swapping does not represent a discontinuity or a transition between independent physical solutions. Rather, it characterizes the position of the physical trajectory within the algebraic root structure. The same analysis also identifies an exact lower bound on the second SR parameter for downhill inflaton motion, with the flat-potential USR limit lying on its boundary. Independently, the conservation of the momentum conjugate to $\calR$ in the long-wavelength limit provides a criterion for the growth or decay of the gauge-invariant curvature difference $|\zeta-\calR|$. This criterion follows from the relation $\zeta-\calR=-d\calR/dN/3$, where $N=\log{a}$ is the number of $e$-folds. The resulting growth boundary is distinct from both the algebraic branch-swap boundary and the limiting boundary of downhill motion.

Combining these results, we construct a unified dynamical chart in the $(\epsilon,\chi)$ plane, where $\chi\equiv\sqrt{\epsilon_V/\epsilon}$. The chart distinguishes the algebraic velocity branches from the large-scale growth or decay of the curvature difference. We further derive a criterion involving the local curvature of the potential that determines the direction of algebraic branch crossing. We illustrate this classification using constant-roll trajectories and simple toy models, demonstrating how different dynamical regimes can be traversed during inflation. More broadly, our framework shifts the emphasis from assigning inflationary models to fixed categories toward understanding the dynamical histories traversed by their trajectories and their implications for primordial perturbations.

This article is organized as follows. In Section~\ref{sec:cubic}, we derive the depressed cubic equation for the inflaton velocity and examine its algebraic root structure. In Section~\ref{sec:decayingR}, we discuss the large-scale evolution of curvature perturbation and the physical interpretation of the curvature-difference growth boundary. In Section~\ref{sec:chart}, we combine these results to construct the dynamical chart in the $(\epsilon,\chi)$ plane. We illustrate its application to simple toy models in Section~\ref{sec:examples} and conclude in Section~\ref{sec:conc}.

\section{Background dynamics: Depressed cubic}
\label{sec:cubic}

The basic background equations for canonical single-field inflation are:
\begin{align}
\label{eq:Friedmann}
H^2 & = \frac{1}{3\mpl^2} \bigg[ \frac{1}{2}\dot\phi^2 + V(\phi) \bigg] \, ,
\\
\label{eq:Hevol}
\dot{H} & = - \frac{\dot\phi^2}{2\mpl^2} \, ,
\\
\label{eq:field-eq}
\ddot\phi + 3H\dot\phi + V_\phi & = 0 \, .
\end{align}
Note that these equations are not all independent. For example, by taking a time derivative of \eqref{eq:Friedmann} and using \eqref{eq:field-eq}, one can immediately derive \eqref{eq:Hevol}. The first SR parameter $\epsilon$ is defined by
\begin{equation}
\label{eq:1st-SR}
\epsilon \equiv -\frac{\dot{H}}{H^2} = \frac{1}{2\mpl^2} \frac{\dot\phi^2}{H^2} \, ,
\end{equation}
where for the second equality we have used \eqref{eq:Hevol}. The second SR parameter, usually denoted by $\eta$, is defined by
\begin{equation}
\label{eq:2nd-SR}
\eta \equiv \frac{\dot\epsilon}{H\epsilon} \, .
\end{equation}
From the second expression of \eqref{eq:1st-SR}, we can rewrite $\eta$ as
\begin{equation}
\label{eq:eta-2nd}
\eta 
= 
\frac{1}{H\epsilon} \frac{1}{\mpl^2} \bigg( \frac{\ddot\phi\dot\phi}{H^2} - \frac{\dot{H}}{H^3}\dot\phi^2 \bigg)
=
- 6 - \frac{2V_\phi}{H\dot\phi} + \frac{1}{\mpl^2} \frac{\dot\phi^2}{H^2}
\, .
\end{equation}
Notice that from the first equality above, we can express $\ddot\phi$ as
\begin{equation}
\label{eq:ddotphi}
\frac{\ddot\phi}{H\dot\phi}
= 
- \epsilon + \frac{\eta}{2}
\, .
\end{equation}
Thus, defining a dimensionless quantity
\begin{equation}
u \equiv \frac{\dot\phi}{\sqrt{2}\mpl H} \, ,
\end{equation}
$u$  always satisfies the following ``depressed cubic'' equation:
\begin{equation}
\label{eq:depressed-cubic}
u^3 - \frac{ 6 + \eta }{2} u - \frac{V_\phi}{\sqrt{2}\mpl H^2} 
\equiv u^3 + A u + B
= 0 \, .
\end{equation}
This general equation is symmetric under the simultaneous change of sign $(u,V_\phi) \to (-u,-V_\phi)$, which relates its two rolling directions.

The nature of the roots of \eqref{eq:depressed-cubic} can be determined without computing them explicitly, by using the discriminant $\Delta$. The discriminant of \eqref{eq:depressed-cubic} is
\begin{equation}
\label{eq:discriminant}
\Delta \equiv \frac{B^2}{4} + \frac{A^3}{27}
\, .
\end{equation}
The discriminant and the nature of the corresponding solutions of \eqref{eq:depressed-cubic}, given that $A$ and $B$ are real numbers, are summarized in Table~\ref{table:discriminant-root}. For $\Delta = 0$ and $A = 0$, we must have $B=0$ too; then the triple root of the cubic is trivially $u=0$. So we do not consider this trivial case. Notice that for $\Delta < 0$, we must have $A<0$ since $B^2 \geq 0$. But this alone does not guarantee the existence of three distinct real roots of \eqref{eq:depressed-cubic}.

\begin{table}[ht!]
\begin{center}
\begin{tabular}{c||l}
Discriminant & Roots
\\\hline\hline
$\Delta < 0$ & All are distinct and real
\\
$\Delta > 0$ & One is real, and the other two are complex and conjugate to each other
\\
$\Delta = 0$ (and $A\neq0$) & A simple root $3B/A$ and a double root $-3B/(2A)$
\end{tabular}
\caption{Different values of the discriminant and the corresponding nature of the roots of \eqref{eq:depressed-cubic}.}
\label{table:discriminant-root}
\end{center}
\end{table}

Now we consider the case $V_\phi > 0$ and $\dot\phi < 0$, i.e. the potential has a positive gradient and the field is rolling from right to left. Although we restrict ourselves to this specific case, other cases can be considered on an equal basis. The purpose of choosing this downhill trajectory is to expose the physical root structure without ambiguity. In this case, $u < 0$ and $B < 0$, so let us redefine $x \equiv -u$, $p \equiv (6+\eta)/2 = -A$ and $q \equiv V_\phi/(\sqrt{2}\mpl H^2) = -B > 0$. Then, \eqref{eq:depressed-cubic} is written as
\begin{equation}
\label{eq:cubic-branch}
x^3 - px + q = 0 \, .
\end{equation} 
This simple relation leads to a number of interesting results. First, we can write $q$ as
\begin{equation}
\label{eq:q}
q = x (p - x^2) \, .
\end{equation}
Since $q>0$ in the current case, we demand $p > x^2$, viz. we constrain the allowed range of $\eta$ as
\begin{equation}
\label{eq:eta-bound}
\eta > -6 + 2\epsilon \equiv \eta_\text{bound} \, .
\end{equation}
Moreover, from \eqref{eq:1st-SR} we already have an {\it exact} physical solution 
\begin{equation}
\label{eq:x0}
x = \sqrt{\epsilon} \equiv x_0 > 0 \, .
\end{equation}
This is consistent with \eqref{eq:q}, which we can trivially check by combining with the last expression in \eqref{eq:eta-2nd}. Thus, we can find that \eqref{eq:cubic-branch} is factorized into the form
\begin{equation}
(x-\sqrt{\epsilon}) \big[ x^2 + \sqrt{\epsilon}x + (\epsilon-p) \big] = 0 \, .
\end{equation}
Thus, other than \eqref{eq:x0}, the quadratic part of the above expression gives the other two roots:
\begin{equation}
\label{eq:xpm}
x
= 
\frac{-\sqrt{\epsilon} \pm \sqrt{2(6+\eta)-3\epsilon}}{2} 
\equiv x_\pm 
\, .
\end{equation}
Since we demand $6 + \eta > 2\epsilon$, equivalently $2(6+\eta)-3\epsilon > \epsilon$, which in turn means $x_+>0$ and $x_-<0$, both real. So two positive roots -- $x_0$ and $x_+$ -- are compatible with the downhill sign. In fact, this can be understood from \eqref{eq:discriminant} very easily as follows. Since the physical root is given by \eqref{eq:x0}, plugging this back into \eqref{eq:q} gives the discriminant of \eqref{eq:cubic-branch} as the following factorized form:
\begin{equation}
\label{eq:discriminant2}
\Delta 
= 
\frac{q^2}{4} - \frac{p^3}{27}
=
\frac{\epsilon(p-\epsilon)^2}{4} - \frac{p^3}{27}
=
\frac{1}{4} \bigg( \epsilon - \frac{4}{3} p \bigg) \bigg( \epsilon - \frac{p}{3} \bigg)^2
\, .
\end{equation}
Since $p > x^2 = \epsilon$, we always have $\epsilon - 4p/3 < 0$ and $(\epsilon-p/3)^2\geq0$. This means $\Delta \leq 0$ always, so there are always three distinct real roots ($\Delta<0$) or at least one simple root and one double root ($\Delta=0$) for a downhill motion of the inflaton.

Also notice from \eqref{eq:eta-bound} it is obvious that, since $\epsilon > 0$, $\eta$ is bounded from below by $\eta_\text{bound}$ for a downhill trajectory. That is, when a canonical single inflaton is monotonically downhill-rolling, $\eta$ cannot be smaller than $\eta_\text{bound}$. Thus $\eta>\eta_\text{bound}$ throughout a strictly downhill phase, with the exact flat-potential boundary given by $\eta=\eta_\text{bound}$. In the quasi-de Sitter limit this approaches $\eta=-6$, showing that USR represents the maximally rapid damping of the kinetic energy compatible with downhill motion~\cite{Kinney:2005vj,Pattison:2018bct}. Thus, USR is distinguished not merely by its familiar non-attractor perturbation dynamics, but also by its position at the limiting boundary of downhill inflaton motion.

Then one may naturally ask. Is any lower value of $\eta$ prohibited? It is not, but the situation is qualitatively different. Let us consider, say, $\eta = -9$ and $\epsilon\ll1$. Then from \eqref{eq:ddotphi}
\begin{equation}
\frac{\ddot\phi}{H\dot\phi} \approx -\frac{9}{2} = -3 - \frac{3}{2} \, ,
\end{equation}
but from \eqref{eq:1st-SR}
\begin{equation}
\frac{\ddot\phi}{H\dot\phi} = -3 - \frac{V_\phi}{H\dot\phi} \, .
\end{equation}
Comparing these two relations, we find 
\begin{equation}
\frac{V_\phi}{H\dot\phi} \approx \frac{3}{2} \, .
\end{equation}
That is, $V_\phi/\dot\phi > 0$, i.e. $V_\phi$ and $\dot\phi$ have the same sign. If $V_\phi > 0$, i.e. potential is rising to the right, $\dot\phi > 0$, viz. the field is moving to the right -- executing an uphill motion. This is physically reasonable. In the downhill motion, the potential force $V_\phi$ is against the Hubble friction $-3H\dot\phi$ to keep $\phi$ rolling, so $\big|\dot\phi\big|$ cannot decrease faster than $a^{-3}$. But in the uphill motion, both $V_\phi$ and $-3H\dot\phi$ act to stop $\phi$, and $\big|\dot\phi\big|$ may decrease faster than $a^{-3}$. Thus, a regime such as $\eta = -9$ is not prohibited, but it does not belong to a standard downhill extension. Usually, such a climbing motion is transient, only until $\dot\phi$ reaches its turning point. At such a point, $\eta = \dot\epsilon/(H\epsilon)$ becomes singular, so on general ground there is no global lower bound for $\eta$ in canonical inflation.

Now we return to the three roots of the depressed cubic equation. As we have seen in \eqref{eq:discriminant2}, the discriminant $\Delta$ can be written in a factorized form, and can be as large as $\Delta = 0$. At this point, the two positive roots -- $x_0$ and $x_+$ -- become identical:
\begin{equation}
x_+ = x_0 \, .
\end{equation}
Plugging this back into \eqref{eq:xpm}, we find that these two roots merge when $\eta$ becomes the following value:
\begin{equation}
\label{eq:eta-boundary}
\eta = -6 (1-\epsilon) \equiv \eta_\text{swap} \, .
\end{equation}
Thus, when $\epsilon \ll 1$, the point of double root is not $\eta = -6$ or $\eta_\text{bound}$, but very close there. 
Equivalently, we can write
\begin{equation}
\frac{\ddot\phi}{H\dot\phi} \bigg|_\text{swap} = - 3 + 2\epsilon \, .
\end{equation}
That is, for $\epsilon\ll1$, the branch swapping occurs near the USR value $-3$, deviating by $\calO(\epsilon)$. Since the true solution is $x_0$, we can summarize the location of $x_0$ depending on the value of $\eta$ as the following: 
\begin{table}[h!]
\begin{center}
\begin{tabular}{c||l}
$\eta > \eta_\text{swap}$ & $x_0 < x_+$, lower-velocity branch 
\\
$\eta = \eta_\text{swap}$ & $x_0=x_+$, double root
\\
$\eta_\text{bound} < \eta < \eta_\text{swap}$ & $x_0>x_+$, higher-velocity branch 
\end{tabular}
\end{center}
\end{table}

\noindent Thus, across $\eta = \eta_\text{swap}$, the physical solution smoothly swaps from lower-velocity branch to higher one. It is then natural to ask what the three roots of the depressed cubic mean. Although all three satisfy the cubic equation \eqref{eq:cubic-branch} for given $(H,V_\phi,\eta)$, not all correspond to physically allowed background solutions. The only possible physical solution is, as mentioned before, $x_0$, imposed by the Friedmann equation \eqref{eq:Friedmann}. What happens across $\eta = \eta_\text{swap}$ is that the physical solution $x_0$ swaps where it is located. This can be viewed as follows. Let us rewrite \eqref{eq:cubic-branch} in such a way that $\eta$ is a function of $x$ and $q$:
\begin{equation}
\eta(x,q) = -6 + 2x^2 + \frac{2q}{x} \, .
\end{equation}
For simplicity, we fix $q$ then $\eta\to\infty$ for both $x\to0$ and $x\to\infty$ (remember that $x > 0$). Thus, there must exist a minimum of $\eta$ at (say) $x_\star$ that can be found as usual:
\begin{equation}
\frac{\partial\eta}{\partial x} \bigg|_{x_\star}
=
4x_\star - \frac{2q}{x_\star^2}
=
0 \, ,
\end{equation}
so that $q - 2x_\star^3 = 0$, or more explicitly
\begin{equation}
\frac{V_\phi}{\sqrt{2}\mpl H^2} + 2\frac{\dot\phi}{\sqrt{2}\mpl H} \epsilon
= 
\frac{\dot\phi}{\sqrt{2}\mpl H} \bigg( -\frac{6+\eta}{2}+3\epsilon \bigg) 
= 
0 \, ,
\end{equation}
which is simply the root-swapping condition \eqref{eq:eta-boundary}. Notice that the branch-swap point is a fold of the inverse algebraic map, rather than a dynamical bifurcation of the inflationary background. The physical trajectory remains regular across this boundary, although its ordering relative to the other positive algebraic root changes.

Thus, we can extract the nature of the three roots in the limit $q\ll1$ in \eqref{eq:cubic-branch}, which is relevant both for SR and USR. First, solving \eqref{eq:cubic-branch} with $q=0$ exactly gives 
\begin{equation}
x^3 - px = x(x^2 - p) = 0 \, ,
\end{equation}
so that the three roots are given by $x = 0$ and $\pm\sqrt{p}$. Now, to take into account the non-zero but small $q$, for each solution we set $x = x_{q=0} + a_1q + \cdots$. 
\begin{enumerate}

\item $x=0$: In this case, simply
\begin{equation}
x = a_1q + \cdots \, ,
\end{equation}
and plugging this into the original equation \eqref{eq:cubic-branch} and only keeping linear order in $q$,
\begin{equation}
x^3 - px + q 
= 0
\approx (-pa_1 + 1)q
\, ,
\end{equation}
so that $a_1 = 1/p$. Thus, in this case
\begin{equation}
x_1 = \frac{q}{p} + \cdots 
\end{equation}
Since both $p$ and $q$ are positive, $x_1>0$.

\item $x=\sqrt{p}$: Similarly, plugging
\begin{equation}
x = \sqrt{p} + a_1q + \cdots
\end{equation}
into \eqref{eq:cubic-branch} and only keeping linear order in $q$ gives
\begin{equation}
\big( p^{3/2} + 3a_1pq + \cdots \big) - p \big( \sqrt{p} + a_1q + \cdots \big) + q
= 0
\approx
\big( 2a_1p + 1 \big) q
\, ,
\end{equation}
so that $a_1 = -1/(2p)$. Thus, in this case
\begin{equation}
x_2 = \sqrt{p} - \frac{q}{2p} + \cdots \, .
\end{equation}
Since we assume $p>0$ and $0<q\ll1$, $x_2>x_1>0$.

\item $x=-\sqrt{p}$: In the same way as $x=\sqrt{p}$, we find
\begin{equation}
x_3 = -\sqrt{p} - \frac{q}{2p} + \cdots < 0 \, .
\end{equation}

\end{enumerate}
Thus, the depressed cubic admits two positive algebraic roots, which we distinguish as the lower- and higher-velocity branches. In the small-$q$ limit, the lower root behaves as $q/p$, while the higher root approaches $\sqrt{p}$. The former contains the conventional SR limit, in which the inflaton velocity is determined primarily by the potential gradient $V_\phi$, whereas the latter contains the flat-potential USR limit, where the inflaton evolves predominantly under the Hubble friction. However, this correspondence holds in the respective limiting regimes and should not be taken as a general identification of the two algebraic branches with SR and USR dynamics. Rather, the branch structure distinguishes the relative ordering of the positive roots, independently of whether the physical inflaton evolution is close to either limiting behaviour. The two positive roots merge at \eqref{eq:eta-boundary} where $\Delta = 0$. We have thus identified two distinct boundaries in the background dynamics -- the branch-swap boundary and the flat-potential USR limit. A separate question is whether the gauge-invariant curvature difference $|\zeta-\calR|$ grows or decays on large scales, which introduces another boundary independent of the background algebraic structure.
Note that the small-$q$ expansions above are valid when $q/p^{3/2}\ll1$, a condition satisfied in both the ordinary SR and deep-USR limits, sufficiently far from the branch-swap point $\Delta=0$. Near this point, however, these expansions break down, and the exact Cardano solutions must be used to describe the merging of the two positive roots. We provide a detailed consistency check in Appendix~\ref{app:cardano}.

\section{Perturbation dynamics: Large-scale evolution}
\label{sec:decayingR}

Having identified the algebraic branch structure of the inflationary background, we now turn to curvature perturbations and seek an additional criterion for their large-scale evolution. A natural question is whether the curvature perturbation approaches a conserved configuration, as in the conventional SR attractor regime, or continues to evolve, as can happen during a USR phase. To address this question, it is useful to consider not only the comoving curvature perturbation $\calR$, but also the curvature perturbation on uniform-density hypersurfaces $\zeta$. In ordinary SR inflation, these two gauge-invariant quantities coincide on sufficiently large scales and become conserved. Outside the SR regime, however, their equivalence is not guaranteed.

For a canonical scalar field in general relativity, an important exact relation connects these two curvature perturbations. In the comoving gauge, the lapse perturbation $A_c$ and the energy-density perturbation $\delta\rho_c$ satisfy
\begin{equation}
\delta\rho_c = - \dot\phi^2 A_c \, ,
\end{equation}
where, from the linear Einstein momentum constraint, we have $A_c = \dot\calR/H$. Since the background energy conservation equation gives $\dot\rho=-3H\dot\phi^2$, the definition of $\zeta$ leads to~\cite{Romano:2015vxz}
\begin{equation}
\label{eq:curvature_difference}
\zeta - \calR
= - H \frac{\delta\rho_c}{\dot\rho}
= - \frac{\dot\calR}{3H}
= - \frac{1}{3} \frac{d\calR}{dN}
\, .
\end{equation}
This relation holds at linear order without requiring the long-wavelength approximation. There is also a geometrical interpretation of this relation. The lapse perturbation on comoving hypersurfaces $A_c=d\calR/dN$ measures the fractional perturbation of their proper-time intervals. Moreover, the infinitesimal time shift from comoving to uniform-density hypersurfaces satisfies
\begin{equation}
\Delta t_{c\to\rho}
= \frac{\delta\rho_c}{\dot\rho}
= \frac{A_c}{3H}
\, ,
\end{equation}
so that $\zeta-\calR=-H\Delta t_{c\to\rho}$. Thus, $\zeta-\calR$ measures the separation between these two physically specified slicings in units of the background expansion time. Importantly, although the time derivative of $\calR$ takes different forms in cosmic time $t$, conformal time $\tau$, and $e$-fold time $N$, the combination
\begin{equation}
\label{eq:curvature_difference_time}
\frac{\dot\calR}{H}
= \frac{1}{aH} \frac{d\calR}{d\tau}
= \frac{d\calR}{dN}
= - 3 ( \zeta - \calR )
\end{equation}
always represents the same gauge-invariant quantity. Thus, the $e$-fold derivative is not singled out merely for convenience: It directly characterizes the difference between the comoving and uniform-density curvature perturbations. This provides a physical motivation for using $N$ to classify their large-scale evolution.

We can further understand the evolution of this difference as follows. The quadratic action for canonical single-field inflation, written in terms of $N$, is
\begin{equation}
\label{eq:action}
S_2 
= 
\mpl^2 \int dN d^3x a^3H\epsilon
\Bigg[ \bigg( \frac{d\calR}{dN} \bigg)^2 - \frac{(\nabla\calR)^2}{a^2H^2} \Bigg] 
\, .
\end{equation}
The corresponding conjugate momentum is
\begin{equation}
\label{eq:PiR}
\Pi_\calR = 2 \mpl^2 a^3 H \epsilon \frac{d\calR}{dN} \, .
\end{equation}
In the strict large-scale limit $k\to0$, we can drop the gradient term and the action depends on $\calR$ only through its derivative, so that the conjugate momentum is conserved:
\begin{equation}
\frac{d\Pi_\calR}{dN} = 0 \, .
\end{equation}
Combining \eqref{eq:curvature_difference} and \eqref{eq:PiR}, we obtain
\begin{equation}
\zeta - \calR
= 
- \frac{1}{3} \frac{d\calR}{dN}
= 
- \frac{\Pi_\calR}{6\mpl^2a^3H\epsilon}
\, .
\end{equation}
This expression reveals the physical mechanism governing the evolution of the curvature difference. Since the conjugate momentum is conserved at $k=0$, a decrease in the positive coefficient $a^3H\epsilon$ necessarily amplifies the magnitude of $\zeta-\calR$, or equivalently $d\calR/dN$, provided the conserved momentum is non-zero. Conversely, an increase in this coefficient suppresses the curvature difference.

The logarithmic rate of change of this coefficient is~\cite{Vallejo-Pena:2019lfo}
\begin{equation}
\label{eq:Gamma_R}
\Gamma_{\calR} 
\equiv
\frac{d\log(a^3H\epsilon)}{dN}
=
3 - \epsilon + \eta
\, .
\end{equation}
Consequently, in the limit $k\to0$,
\begin{equation}
\label{eq:curvature_difference_growth}
\frac{d}{dN} \log|\zeta-\calR| = - \Gamma_\calR \, .
\end{equation}
We can therefore identify a boundary separating the instantaneous amplification and damping of $|\zeta-\calR|$ as $\Gamma_\calR = 0$, or equivalently
\begin{equation}
\eta_\text{grow} \equiv - 3 + \epsilon \, .
\end{equation}
More explicitly, if $\eta < \eta_\text{grow}$ ($\eta > \eta_\text{grow}$), $|\zeta - \calR|$ grows (decays). If the conserved momentum vanishes, the difference remains zero, and no amplification occurs. This gives the boundary $\eta_\text{grow}$ a physical meaning that is not tied to an arbitrary choice of time coordinate. It characterizes the evolution of the difference between two gauge-invariant curvature perturbations in the super-horizon limit. In particular, whether $|\zeta-\calR|$ increases or decreases is unchanged under a monotonic reparametrization of time. The notation $\eta_\text{grow}$ will be used for the boundary for the growth of this curvature difference, rather than as a universal criterion for the growth of the curvature perturbation itself, as we will discuss later.

To make contact with the conventional description, the equation of motion for the Fourier mode $\calR_k$ obtained from \eqref{eq:action} is
\begin{equation}
\label{eq:Req}
\frac{d^2\calR_k}{dN^2} 
+ \underbrace{ ( 3 - \epsilon + \eta ) }_{= \Gamma_\calR} \frac{d\calR_k}{dN}
+ \frac{k^2}{a^2H^2} \calR_k
= 0
\, .
\end{equation}
Thus, $\Gamma_\calR$ also appears as the friction coefficient when the equation for $\calR_k$  is expressed in $N$. The regime $\Gamma_\calR<0$ is commonly associated with the growth of the would-be decaying solution in non-attractor inflation. From the conserved-momentum perspective, however, the precise interpretation is that the decreasing kinetic coefficient $a^3H\epsilon$ amplifies $d\calR/dN$ and, equivalently, the curvature difference $\zeta-\calR$ in the limit $k\to0$. It is worth distinguishing this interpretation from the friction coefficients obtained using other time variables. The conserved momentum itself can equivalently be written as
\begin{equation}
\Pi_{\calR}
= 2\mpl^2a^3\epsilon \dot{\calR}
= 2\mpl^2a^2\epsilon \frac{d\calR}{d\tau}
= 2\mpl^2a^3H\epsilon \frac{d\calR}{dN}
\, .
\end{equation}
Accordingly, in the limit $k\to0$, the instantaneous logarithmic growth rates of the corresponding derivative magnitudes are
\begin{align}
\frac{d}{dN} \Big( \log \big| \dot\calR \big| \Big)
& = 
- 3 - \eta 
\, ,
\\
\frac{d}{dN} \bigg( \log \bigg| \frac{d\calR}{d\tau} \bigg| \bigg)
&=
- 2 - \eta
\, ,
\\
\frac{d}{dN} \bigg( \log \bigg| \frac{d\calR}{dN} \bigg| \bigg)
& =
- 3 + \epsilon - \eta
\, .
\end{align}
The three conditions for the growth of the time derivative of $\calR$ are therefore different. Nevertheless, these expressions do not describe conflicting physical predictions. They concern derivatives defined with respect to different time coordinates. The combination $\dot{\calR}/H=d\calR/dN$ is distinguished for our purposes because of its direct relation to $\zeta-\calR$ in \eqref{eq:curvature_difference_time}. Hence $\Gamma_\calR$ is not a coordinate-independent friction coefficient, but it does characterize the coordinate-independent growth of a specified gauge-invariant curvature difference.

An important qualification is that the growth of $|\zeta-\calR|$ does not necessarily imply enhancement of the complete curvature amplitude $|\calR|$. To see this, we consider the general large-scale solution of \eqref{eq:Req}: 
\begin{equation}
\calR(N) 
=
C_1 + C_2 \int_{N_i}^N \frac{dN'}{a^3H\epsilon}
\equiv 
C_1 + C_2 I(N)
\, .
\end{equation}
The first term is the constant solution, whereas the second is the non-constant solution, conventionally referred to as the would-be decaying mode. However, an individual non-constant solution is defined only up to the addition of a constant, since $I(N)+A$ is an equally valid non-constant solution for any constant $A$. Therefore, whether the magnitude of a particular non-constant solution grows or decays requires a specified normalization or asymptotic definition. It cannot generally be determined by the instantaneous value of $\Gamma_\calR$ alone. By contrast, the complete curvature perturbation and the curvature difference $\zeta-\calR$ are unaffected by this ambiguity: The latter depends only on the derivative of $\calR$, which is insensitive to an arbitrary addition of the constant solution. The amplitude of the complete curvature perturbation satisfies
\begin{equation}
\frac{d|\calR|}{dN}
=
\frac{1}{|\calR|} \Re \bigg( \calR^* \frac{d\calR}{dN} \bigg)
\, .
\end{equation}
Thus the evolution of the complete amplitude depends on the relative magnitudes and phases of the two independent contributions. The background condition $\Gamma_\calR<0$ alone does not determine the instantaneous growth of $|\calR|$ or its final power spectrum. The coefficients $C_1$ and $C_2$ are fixed by the specified initial state and the preceding background evolution. For a time-dependent background, the integrated evolution makes this distinction particularly transparent. Integrating \eqref{eq:curvature_difference_growth} from $N_i$ to $N_f$ gives
\begin{equation}
\log \frac{\big| \zeta - \calR \big|_{N_f}}{\big| \zeta - \calR \big|_{N_i}}
=
- \int_{N_i}^{N_f} \Gamma_\calR(N) dN
\, .
\end{equation}
Meanwhile, the curvature perturbation itself evolves according to its equation of motion, and in the limit $k\to0$  integrating \eqref{eq:Req} twice from $N_i$ to $N_f$ gives
\begin{equation}
\calR(N_f)
=
\calR(N_i) + \frac{d\calR}{dN}(N_i) 
\int_{N_i}^{N_f} dN \exp \bigg[ - \int_{N_i}^N \Gamma_\calR(N') dN' \bigg]
\, .
\end{equation}
Thus, the growth of the curvature difference is determined by the accumulated evolution of $\Gamma_\calR$, whereas the amplitude of the complete curvature perturbation additionally depends on the initial curvature perturbation, its derivative, and their relative phase. These relations clarify why a local dynamical boundary cannot by itself predict the final power spectrum.

Finally, we emphasize the domain of validity of this classification. The relation $\zeta-\calR=-d\calR/dN/3$ holds at linear order for a canonical scalar field in general relativity, but the conservation of $\Pi_\calR$ is exact only at $k=0$. At finite wavenumber, the evolution equation gives [see \eqref{eq:curvature_difference_growth} in the $k\to0$ limit]
\begin{equation}
\frac{d}{dN} \log \big| \zeta_k - \calR_k \big|
=
- \Gamma_\calR 
+ \frac{k^2}{3a^2H^2} \frac{\Re \big[ \big( \zeta_k^* - \calR_k^* \big) \calR_k \big]}{\big| \zeta_k - \calR_k \big|^2}
\, .
\end{equation}
Therefore, even when $k\ll aH$, the gradient contribution may become important if $\zeta_k - \calR_k$ or $\Gamma_\calR$ is sufficiently small. In particular, the background boundary $\Gamma_\calR=0$ should be understood as the leading-order growth boundary, rather than an exact boundary for arbitrary finite-wavelength modes.

We have thus identified a perturbative boundary with a well-defined physical interpretation. It distinguishes the growth and decay of the difference between comoving and uniform-density curvature perturbations at leading order in the large-scale limit. This boundary follows from the momentum conservation and is distinct from the algebraic velocity-branch structure of the background. Combining these two structures allows us to construct a dynamical chart of canonical single-field inflation.

\section{A dynamical chart of single-field inflation}
\label{sec:chart}

Combining the algebraic background structure derived in Section~\ref{sec:cubic} with the large-scale curvature-difference evolution discussed in Section~\ref{sec:decayingR}, we can now classify canonical single-field inflationary dynamics according to two complementary criteria: The ordering of the positive roots for the inflaton velocity and the growth or decay of $|\zeta-\calR|$. To present this classification more transparently, let us introduce the potential SR parameter:
\begin{equation}
\epsilon_V \equiv \frac{\mpl^2}{2} \bigg( \frac{V_\phi}{V} \bigg)^2 \, .
\end{equation}
Then, we can write
\begin{equation}
\label{eq:V'-eta}
\frac{V_\phi}{H\dot\phi}
= 
\frac{\sqrt{2\epsilon_V}}{\mpl} \frac{V}{H\dot\phi}
=
- (3-\epsilon) \chi 
\, ,
\end{equation}
where we have defined
\begin{equation}
\chi \equiv \sqrt{\frac{\epsilon_V}{\epsilon}} \, .
\end{equation}
Importantly, the relative ordering of the curvature-difference growth and branch-swap boundaries depends on $\epsilon$, since $\eta_\text{grow}-\eta_\text{swap}=3-5\epsilon$. For $\epsilon<3/5$, the curvature difference begins to grow before the physical solution reaches the higher-velocity branch as $\eta$ decreases at fixed $\epsilon$. For $3/5<\epsilon<1$, the ordering is reversed, allowing a regime in which the physical solution occupies the higher-velocity branch while $|\zeta-\calR|$ decreases. Thus, the background algebraic branch structure and the large-scale evolution of the gauge-invariant curvature difference provide two complementary criteria for classifying inflationary dynamics.

\begin{table}[h!]
\begin{center}
\begin{tabular}{llcc}
Branch & $|\zeta-\calR|$ & Conditions & Allowed $\epsilon$
\\\hline\hline
\multirow{2}{*}{Lower velocity} & \multirow{2}{*}{Decaying} & 
$\eta > \max(\eta_\text{swap},\eta_\text{grow})$ & \multirow{2}{*}{$0<\epsilon<1$}
\\
& & $\chi > \max(\chi_\text{swap},1/2)$ &
\\\hline
\multirow{2}{*}{Lower velocity} & \multirow{2}{*}{Growing} 
& $\eta_\text{swap} < \eta < \eta_\text{grow}$ & \multirow{2}{*}{$0<\epsilon<\dfrac{3}{5}$}
\\
& & $\chi_\text{swap} < \chi < 1/2$ &
\\\hline
\multirow{2}{*}{Higher velocity} & \multirow{2}{*}{Decaying} & 
$\eta_\text{grow} < \eta < \eta_\text{swap}$ & \multirow{2}{*}{$\dfrac{3}{5}<\epsilon<1$}
\\
& & $1/2 < \chi < \chi_\text{swap}$ &
\\\hline
\multirow{2}{*}{Higher velocity} & \multirow{2}{*}{Growing} & 
$\eta_\text{bound} < \eta < \min(\eta_\text{swap},\eta_\text{grow})$ & \multirow{2}{*}{$0<\epsilon<1$}
\\
& & $0 < \chi < \min(\chi_\text{swap},1/2)$ &
\end{tabular}
\caption{Classification of downhill canonical single-field inflation according to the algebraic velocity branch and the super-horizon growth or decay of $|\zeta-\calR|$. The branch-swap and curvature-difference growth boundaries are respectively defined by $\eta_\text{swap}=-6(1-\epsilon)$ and $\eta_\text{grow}=-3+\epsilon$, with the downhill bound $\eta_\text{bound}=-6+2\epsilon$. The two positive roots merge at $\eta=\eta_\text{swap}$, whereas the logarithmic growth rate of a non-zero $|\zeta-\calR|$ vanishes at $\eta=\eta_\text{grow}$. The two boundaries intersect at $\epsilon=3/5$.}
\label{table:chart}
\end{center}
\end{table}

Note that here to fix the sign convention we have assumed a monotonic potential with $\phi$ rolling towards the origin from the right, i.e. $V_\phi>0$ and $\dot\phi < 0$. But the sign convention can be chosen on case-basis and does not harm physics. Plugging \eqref{eq:V'-eta} into \eqref{eq:eta-2nd} we can write
\begin{equation}
\label{eq:eta-chi}
\eta
=
2 (3-\epsilon) ( \chi - 1 )
\, .
\end{equation}
In particular, combining this relation with \eqref{eq:Gamma_R}, we obtain
\begin{equation}
\Gamma_\calR = (3-\epsilon)(2\chi-1) \, ,
\end{equation}
so that the curvature-difference growth boundary $\eta=\eta_\text{grow}$ corresponds simply to $\chi=\chi_\text{grow}=1/2$, independently of $\epsilon$ within the inflationary regime.
The four possible combinations of algebraic branch and the behaviour of $|\zeta-\calR|$ are summarized in Table~\ref{table:chart}, with their ordering depending on whether $\epsilon$ is smaller or larger than $3/5$. Table~\ref{table:chi-classification} presents the individual criteria for distinguishing these regimes in the $(\epsilon,\chi)$ plane.

\begin{table}[h!]
\begin{center}
\begin{tabular}{l|l|l}
\multirow{2}{*}{Branch} & $\chi > \chi_\text{swap} = \dfrac{2\epsilon}{3-\epsilon}$ & Lower-velocity branch
\\\cline{2-3}
& $0 < \chi < \chi_\text{swap}$ & Higher-velocity branch
\\\hline
\multirow{2}{*}{$|\zeta-\calR|$} & $\chi > \chi_\text{grow} = \dfrac{1}{2}$ & $|\zeta-\calR|$ decays
\\\cline{2-3}
& $0 < \chi < \chi_\text{grow}$ & $|\zeta-\calR|$ grows
\\\hline\hline
Velocity reference & $\chi = 1$ & $\eta=0$
\\
Downhill boundary & $\chi = 0$ & $V_\phi = 0$
\end{tabular}
\caption{Dynamical classification of downhill canonical single-field inflation in the ($\epsilon$,$\chi$) plane. The velocity branch and the large-scale growth or decay of $|\zeta-\calR|$ are characterized independently by $\chi_\text{swap}$ and $\chi_\text{grow}$ respectively. The last two rows indicate the reference line and the flat-potential boundary.} 
\label{table:chi-classification}
\end{center}
\end{table}

Here, let us pause and consider the physical meaning of $\chi$ more carefully. If we invert \eqref{eq:eta-chi}
\begin{equation}
\chi = 1 + \frac{\eta}{2(3-\epsilon)} \, .
\end{equation}
Thus, if $\epsilon$ increases as inflation proceeds, i.e. $\eta = d\log\epsilon/dN > 0$, then $\chi > 1$. Furthermore, as the SR velocity is written as
\begin{equation}
\dot\phi_\text{SR} = - \frac{V_\phi}{3H} \, ,
\end{equation}
we can write an exact relation of $\chi$ as
\begin{equation}
\chi 
= 
\frac{3}{3-\epsilon} \frac{-\dot\phi_\text{SR}}{\dot\phi}
\approx 
\frac{\big|\dot\phi_\text{SR}\big|}{\big|\dot\phi\big|} 
\, ,
\end{equation}
where the last approximation holds as long as $\epsilon\ll1$. 
Thus, $\chi$ measures the mismatch between the actual inflaton velocity and its SR reference value. It also determines the instantaneous inflaton acceleration through $\ddot\phi/(H\dot\phi)=-3+(3-\epsilon)\chi$. Thus, sufficiently small $\chi$ corresponds to USR-like evolution, while values near unity correspond to the usual SR regime in the quasi-de Sitter limit. Importantly, however, these dynamical descriptions should not be identified with the lower- and higher-velocity branches of the depressed cubic. Indeed, for $\epsilon\ll1$, the branch-swap boundary lies at $\chi_{\mathrm{swap}}\approx2\epsilon/3$, so that a trajectory can exhibit nearly USR-like acceleration while remaining on the lower-velocity branch. The algebraic root structure and the actual rate of roll therefore provide complementary characterizations of the background dynamics. During the downhill motion, as we have set $\dot\phi < 0$ and $V_\phi > 0$, in principle $\chi$ may span a wide range of possible values, $0\leq\chi<\infty$. Thus, we may summarize each regime of $\chi$ as follows:
\begin{itemize}

\item $\chi=0$: In this case, $\epsilon_V = 0$, which means $V_\phi = 0$. Notice that $V_\phi = 0$ does not immediately mean a USR phase. A sustained USR phase additionally requires this condition to persist over a sufficiently long interval. Remember that the lower bound of $\eta$ for the downhill motion is $\eta_\text{bound} = - 6 + 2\epsilon$.

\item $0 < \chi < \chi_\text{swap}$: In this regime, the physical solution occupies the higher-velocity branch. For $\epsilon\ll1$, this region is also close to the USR limit.

\item $\chi_\text{swap} < \chi < 1/2$: This is an interesting regime, where now the physical solution remains on the lower-velocity branch, although $|\zeta-\calR|)$ grows. Notice that this regime exists only for $\epsilon<3/5$.

\item $1/2 < \chi < 1$: In this regime, $|\zeta-\calR|$ decays and $\eta<0$, so that $\epsilon$ decreases. However, the physical solution may lie on either the lower- or higher-velocity branch, depending on its position relative to $\chi_\text{swap}$. The latter possibility arises only for $\epsilon>3/5$. Thus, the actual inflaton velocity remains larger in magnitude than its SR reference value.

\item $\chi = 1$: This is the exact $\eta=0$ reference line, corresponding to constant $\epsilon$ and $\epsilon_V=\epsilon$. In the quasi-de Sitter limit, it approaches the conventional SR balance, although $\ddot\phi/(H\dot\phi)=-\epsilon$ rather than exactly zero.

\item $\chi > 1$: In this regime, $\eta>0$, so $\epsilon$ increases as inflation proceeds. However, this does not necessarily imply that the magnitude of the inflaton velocity increases. The field accelerates along its downhill direction only for $\chi>3/(3-\epsilon)$, which is also the condition under which the actual velocity becomes smaller in magnitude than its local SR reference value. In the quasi-de Sitter limit, this threshold approaches $\chi=1$.

\end{itemize}

\begin{figure}[ht!]
\begin{center}
\includegraphics[width=0.48\textwidth]{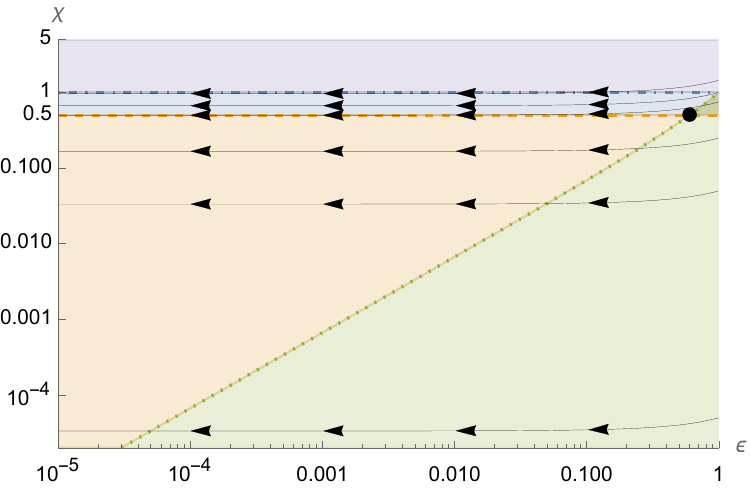}
\includegraphics[width=0.48\textwidth]{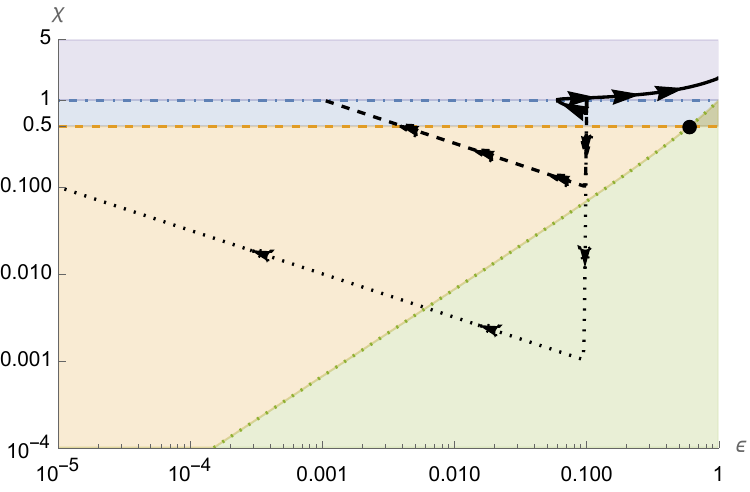}
\end{center}
\caption{Regimes of distinctive inflationary dynamics are presented in the ($\epsilon,\chi$) plane. In both panels, we indicate the SR reference line ($\chi=1$, dot-dashed), the curvature-difference growth boundary ($\chi=1/2$, dashed), and the branch-swap boundary [$\chi=2\epsilon/(3-\epsilon)$, dotted]. The shaded regions correspond to the dynamical regimes classified in Table~\ref{table:chart}. The dot in the upper right region denotes the point $(\epsilon,\chi)=(3/5,1/2)$ where the two boundaries -- $\chi_\text{grow} = 1/2$ and $\chi_\text{swap}$ -- intersect. Also shown are (left panel) families of constant-roll trajectories with, from top to bottom, $\beta=-0.1$, $-1$, $-1.5$, $-2.5$, $-2.9$ and $-2.9999$, and (right panel) the trajectories for the smooth slope-transition model given by \eqref{eq:tanh-potential}, starting from $(\epsilon,\chi) = (0.1,1.0)$, for $\rho = 0.7$ (solid), $0.1$ (dashed) and $0.001$ (dotted). Arrows indicate the direction of evolution with increasing $e$-fold number $N$.}
\label{fig:epsilon-chi}
\end{figure}

\noindent In Figure~\ref{fig:epsilon-chi} we show the chart for different dynamical regimes in the ($\epsilon$,$\chi$) plane. Notice that while the chart provides an exact instantaneous classification, the trajectory through it depends on additional information about the potential, in particular its curvature. That is, with additional potential SR parameter $\eta_V$ defined by
\begin{equation}
\eta_V \equiv \mpl^2 \frac{V_{\phi\phi}}{V} \, ,
\end{equation}
$\epsilon$ and $\chi$ evolve during downhill motion respectively according to 
\begin{align}
\label{eq:dedN}
\frac{d\epsilon}{dN} & = 2\epsilon (3 - \epsilon) (\chi - 1) \, ,
\\
\label{eq:dchidN}
\frac{d\chi}{dN} & = -\eta_V + (3 - \epsilon) \chi - 3 (1 - \epsilon) \chi^2 \, .
\end{align}
These evolution equations apply within the monotonic downhill motion, away from turning points and stationary points of the potential, with the boundaries understood by continuity where appropriate. Thus, to determine which trajectory is actually followed, we need the curvature of the potential via $\eta_V$.

Let us define
\begin{equation}
F(\epsilon,\chi) 
\equiv 
\chi - \chi_\text{swap}(\epsilon) 
= 
\chi - \frac{2\epsilon}{3-\epsilon}
\, .
\end{equation}
Obviously, $F > 0$ ($F<0$) corresponds to the lower-velocity (higher-velocity) branch. Now, differentiating $F$ with respect to $N$ and using \eqref{eq:dedN} and \eqref{eq:dchidN}, we find at $F = 0$, i.e. at the point of root swapping, the following exact relation:
\begin{equation}
\label{eq:crossing-criterion}
\frac{dF}{dN} \bigg|_\text{swap}
= 
- \eta_V + \frac{2\epsilon (9 - 7\epsilon)}{3-\epsilon}
\, .
\end{equation}
Thus, if $\eta_V > 2\epsilon (9 - 7\epsilon)/(3-\epsilon)$, the trajectory proceeds from the lower-velocity branch to the higher one. If the inequality sign is opposite, the trajectory goes the other way around. Thus, the local direction of its algebraic branch transition can be determined directly from the curvature of the potential.

More importantly, this illustrates that our classification goes beyond a static division of the parameter space. Together with the evolution equations \eqref{eq:dedN} and \eqref{eq:dchidN}, the crossing criterion \eqref{eq:crossing-criterion} connects the properties of the potential to the direction of transitions between algebraic velocity branches, while trajectories in the $(\epsilon,\chi)$ plane reveal which dynamical regimes are traversed, in what order, and for how long. Thus, the chart provides a framework for organizing inflationary histories, rather than merely restating known relations among SR parameters.

Before we proceed, we point out that our dynamical classification naturally accommodates constant-roll inflation, defined by a constant $\beta\equiv\ddot\phi/(H\dot\phi)$~\cite{Motohashi:2014ppa}. From \eqref{eq:ddotphi}, $\eta = 2(\beta+\epsilon)$ so that constant-roll solutions follow a one-parameter family of trajectories in the $(\epsilon,\chi)$ plane, given exactly by $\chi=(3+\beta)/(3-\epsilon)$ for downhill motion. In the quasi-de Sitter limit, these trajectories become approximately horizontal, $\chi\approx1+\beta/3$, interpolating between ordinary SR ($\beta\approx0$) and the exact flat-potential USR limit $(\beta=-3)$, as illustrated in the left panel of Figure~\ref{fig:epsilon-chi}. Interestingly, a fixed value of $\beta$ does not necessarily select a unique algebraic velocity branch. The branch-swap line visible in Figure~\ref{fig:epsilon-chi} is crossed at $\epsilon=(3+\beta)/2$, whereas the growth of the curvature difference $|\zeta-\calR|$ is determined independently, at leading order in the large-scale limit, by $\Gamma_\calR=3+\epsilon+2\beta<0$. For instance, the trajectory with $\beta=-5/2$ crosses the branch-swap line at $\epsilon=1/4$ while $|\zeta-\calR|$ continues to grow. Thus, constant-roll inflation represents a restricted family of trajectories within our more general dynamical chart, which distinguishes background branch structure from super-horizon perturbation evolution.

\section{Toy models}
\label{sec:examples}

As a simple toy model that can probe different regions of our dynamical chart, let us consider the following Starobinsky model~\cite{Starobinsky:1992ts}, featuring a linear potential with a sudden change of slope at $\phi = \phi_0$:
\begin{equation}
\label{eq:toymodel}
V_\phi = \left\{
\begin{array}{ll}
s_1 & (\phi > \phi_0)
\vspace{0.3em}\\
s_2 \equiv \rho s_1 & (\phi < \phi_0)
\end{array}
\right.
\, ,
\end{equation}
with $s_1 > s_2 > 0$ (that is, $\rho < 1$) so that $\phi$ rolls from right to left. Assuming SR for $\phi > \phi_0$, we can approximate $\dot\phi \approx -s_1/(3H_0)$ and $\chi \approx 1$. Now, after $\phi$ crosses $\phi_0$, we assume that both $\dot\phi$ and $H$ change negligibly, so that $\epsilon$ remains nearly the same, say, $\epsilon \approx \epsilon_*$. Meanwhile, $\epsilon_V \propto V_\phi^2$ is reduced by $\rho^2$, so that the value of $\chi$ changes from 1 to $\rho$ across $\phi = \phi_0$: 
\begin{equation}
\chi = \left\{
\begin{array}{ll}
1 & (\phi > \phi_0)
\vspace{0.3em}\\
\rho & (\phi < \phi_0)
\end{array}
\right.
\, .
\end{equation}
Thus, depending on the slope ratio $\rho$, the trajectory immediately after the transition can enter any of the following regimes for $\epsilon_*<3/5$:
\begin{table}[h!]
\begin{center}
\begin{tabular}{c||l}
$\rho > \dfrac{1}{2}$ & Lower-velocity branch, $|\zeta-\calR|$ decays
\\
$\dfrac{2\epsilon_*}{3-\epsilon_*} < \rho < \dfrac{1}{2}$ & Lower-velocity branch, $|\zeta-\calR|$ grows
\\
$0 < \rho < \dfrac{2\epsilon_*}{3-\epsilon_*}$ & Higher-velocity branch, $|\zeta-\calR|$ grows
\end{tabular}
\end{center}
\end{table}

\noindent Here and below, the growth or decay of $|\zeta-\calR|$ refers to the leading-order behaviour in the large-scale limit established in Section~\ref{sec:decayingR}.

Moreover, we can solve analytically for $\phi$ for $\phi < \phi_0$ as follows. Right after $\phi$ crosses $\phi_0$, let us write the initial velocity $\dot\phi\big|_\text{after} \equiv - v_*$. Meanwhile, the potential is linear with the slope $s_2$ and $H \approx H_0$. So the velocity approaches $\dot\phi \to -s_2/(3H_0) \equiv -v_\text{SR2}$ at late times. With these boundary conditions, we can solve for $\dot\phi \equiv -v$:
\begin{equation}
\ddot\phi + 3H_0\dot\phi + s_2 
=
- \dot{v} - 3H_0v + s_2
=
- H_0 \bigg[ \frac{d}{dN} \big( v - v_\text{SR2} \big) + 3 \big( v - v_\text{SR2} \big) \bigg]
= 0
\, ,
\end{equation}
where for the last expression we have used the constancy of $v_\text{SR2}$. Thus, setting $N = 0$ initially, the solution that satisfies $v(N = 0) = v_*$ and $v(N\to\infty) = v_\text{SR2}$ is
\begin{equation}
v(N) = v_\text{SR2} + \big( v_* - v_\text{SR2} \big) e^{-3N} \, .
\end{equation}
If initially $v_* \gg v_\text{SR2}$, immediately after $\phi = \phi_0$, $v$ retains its previous value $v_*$ that decays as $v \propto e^{-3N}$, i.e. a USR-like transition occurs. But later $v$ relaxes to its new SR value $v_\text{SR2}$. In terms of $\eta$,
\begin{equation}
\eta
=
- 6 - \frac{2V_\phi}{H\dot\phi} + \frac{1}{\mpl^2} \frac{\dot\phi^2}{H^2}
\approx
- 6 + \frac{2s_2}{H_0v}
=
- 6 \bigg( 1 - \frac{v_\text{SR2}}{v} \bigg)
\, .
\end{equation}
Thus, $\eta$ evolves from approximately $-6(1-\rho)$ immediately after the transition towards zero as the velocity relaxes to its new SR value. In particular, for $\rho\ll1$, the initial evolution is close to the USR limit $\eta\approx-6$. Note that to model a smooth transition from SR to USR (and then back to SR), instead of the piecewise constant values for $V_\phi$ as given in \eqref{eq:toymodel}, a smooth function, e.g. hyperbolic tangent, may be adopted.

In the right panel of Figure~\ref{fig:epsilon-chi}, we present the trajectories for the smooth potential derivative
\begin{equation}
\label{eq:tanh-potential}
V_\phi = \frac{s_1+s_2}{2} + \frac{s_1-s_2}{2} \tanh \bigg( \frac{\phi - \phi_0}{\delta} \bigg) \, ,
\end{equation}
where $\delta$ controls the width of the transition, and $s_2 \equiv \rho s_1$ with $\rho <1$. We set $\delta = 10^{-4}\mpl$ and $\phi - \phi_0 = 0.01\mpl$, and initially the evolution begins at $(\epsilon,\chi) = (0.1,1.0)$. Meanwhile, $V(\phi_0)=10^{-10}\mpl^4$ is chosen with $s_1 \approx 0.449 V(\phi_0)/\mpl$ almost independent of $\rho$. Different trajectories denote different values of $\rho$: $\rho = 0.7$ (solid), $0.1$ (dashed) and $0.001$ (dotted). For $\rho = 0.7$, the slope flattens weakly, so $\chi$ drops only slightly but soon $\dot\phi$ moves more slowly than $\dot\phi_\text{SR}$, and the trajectory enters the regime $\chi>1$. Meanwhile, for $\rho = 0.1$, the trajectory goes below $\chi=1/2$, so that the curvature difference $|\zeta-\calR|$ grows, without crossing $\chi_\text{swap}$. Thus, the trajectory enters the curvature-difference growing regime while remaining on the lower-velocity branch. For $\rho = 0.001$, $\chi$ decreases to cross $\chi_\text{swap}$, entering the higher-velocity branch, which approaches the USR limit for sufficiently small $\epsilon$. Thus, depending on the value of $\rho$, this single model can give rise to a representative family of models that probes different dynamical regimes.

Let us consider another toy model. A particularly interesting regime in our classification is one in which the physical velocity occupies the lower-velocity branch while $|\zeta-\calR|$ grows. As emphasized above, this does not necessarily imply SR background dynamics. To illustrate this regime explicitly, let us consider a quadratic correction to a constant potential:
\begin{equation}
V(\phi) = V_0 + \frac{1}{2} m^2\phi^2 \, .
\end{equation}
If $V_0$ dominates so that $H \approx \sqrt{V_0/(3\mpl^2)} \equiv H_0$, we can find $\phi$ analytically by assuming $\phi \propto e^{rt}$:
\begin{equation}
\ddot\phi + 3H_0\dot\phi + m^2\phi 
=
\big( r^2 + 3H_0r + m^2 \big) \phi = 0
\, .
\end{equation}
Thus we find, for $0<m^2/H_0^2<9/4$, two solutions for the exponent $r$:
\begin{equation}
r = \frac{-3H_0 \pm \sqrt{9H_0^2 - 4m^2}}{2}
= H_0 \bigg( - \frac{3}{2} \pm \sqrt{ \frac{9}{4} - \frac{m^2}{H_0^2}} \bigg)
\equiv 
r_\pm
\, .
\end{equation}
Then, the slow-roll parameter $\epsilon$ evolves as
\begin{equation}
\epsilon \propto e^{2r_\pm t} \, ,
\end{equation}
so that
\begin{equation}
\eta = \frac{2r_\pm}{H_0} = - 3 \pm \sqrt{9 - \frac{4m^2}{H_0^2}} \equiv \eta_\pm
\, .
\end{equation}
For each pure mode that only carries either $r_+$ or $r_-$, the corresponding value of $\eta$ is $\eta_+ > -3$ and $\eta_- < -3$, with $\eta_++\eta_-=-6$, exhibiting a duality structure reminiscent of that discussed in~\cite{Wands:1998yp}. 
Thus, this very simple quadratic plateau provides a dual pair of solutions exhibiting opposite behaviours of the curvature difference: one decaying and the other growing. For example, when $m^2/H_0^2=2$, we obtain $\eta_+ = -2$ and $\eta_- = -4$. In the quasi-de Sitter limit, both solutions occupy the lower-velocity branch, although $|\zeta-\calR|$ decays for the former and grows for the latter.

\section{Conclusions}
\label{sec:conc}

In this work, we have developed an algebraic classification of canonical single-field inflationary dynamics based on an exact depressed cubic equation for the inflaton velocity. Rather than characterizing inflationary models solely through their potential shapes or SR parameters, our approach organizes inflationary evolution according to the algebraic structure of the background equations and the large-scale behaviour of curvature perturbations. We have identified three distinct boundaries: The merging of the algebraic velocity branches, the limiting boundary of downhill inflaton motion, and the onset of growth of the gauge-invariant curvature difference $|\zeta-\calR|$. In particular, the latter boundary has a physical interpretation through the conserved momentum conjugate to $\calR$ and the geometrical separation between comoving and uniform-density hypersurfaces. The significance of our classification lies in distinguishing the algebraic background structure from the perturbation dynamics, while relating the direction of  branch crossing to the local curvature of the potential.

A particularly interesting consequence is that the familiar distinction between SR and USR does not capture the full range of possible dynamics. The lower- and higher-velocity branches should not be identified directly with SR and USR evolution. In the quasi-de Sitter regime, the curvature-difference growth boundary lies well above the branch-swap boundary, leaving a broad region in which the physical velocity occupies the lower-velocity branch while $|\zeta-\calR|$ already grows. Such a background can even exhibit nearly USR-like acceleration without reaching the higher-velocity branch. Conversely, branch swapping does not necessarily coincide with the onset of curvature-difference growth. Our analysis also clarifies the distinguished role of USR. The exactly flat-potential limit represents the boundary of downhill canonical dynamics, rather than a unique condition for large-scale curvature-difference growth. These observations place SR, USR, and their intermediate regimes within a common dynamical framework.

The classification becomes especially transparent in the $(\epsilon,\chi)$ plane. In this representation, different inflationary phases correspond to well-defined regions, while their transitions can be followed as trajectories across the dynamical chart. We have illustrated this picture in Figure~\ref{fig:epsilon-chi} with constant-roll solutions and simple toy potentials, demonstrating that even a single model can explore qualitatively different regimes depending on its parameters and initial conditions. In this sense, our scheme classifies not merely inflationary models, but the dynamical histories realized by those models.

More broadly, our perspective offers a useful starting point for connecting inflationary model building with the evolution of primordial perturbations. Although the growth of $|\zeta-\calR|$ does not by itself guarantee an enhancement of the final curvature power spectrum, the dynamical chart identifies distinct background regimes and characterizes the associated large-scale evolution of the curvature difference. A systematic exploration of trajectories through this chart may therefore help clarify how different inflationary histories influence primordial power spectra, non-Gaussianity, and primordial black hole production. For inflationary model building, the question is not merely where a trajectory lies on the chart, but which dynamical regimes a given potential allows it to traverse, in what order, and for how long. It is this dynamical history, together with the initial perturbation state, that connects the inflationary background to the primordial fluctuations it ultimately produces.

\subsection*{Acknowledgements}

We thank Daiki Saito for helpful discussions.
We are supported in part by the National Research Foundation of Korea Grant (RS-2024-00336507). JG is further supported by the Ewha Womans University Research Grant of 2026 (1-2026-0486-001-1).
JG is grateful to the Asia Pacific Center for Theoretical Physics for hospitality while this work was under progress.

\appendix
\renewcommand{\theequation}{\thesection.\arabic{equation}}
\setcounter{equation}{0}

\section{Cardano solutions and the small-$q$ expansion validity}
\label{app:cardano}

Here we examine the validity of the small-$q$ expansions used in Section~\ref{sec:cubic} and clarify their relation to the exact Cardano solutions, particularly near the branch-swap point. The expansion is performed at fixed $p$ and characterizes the algebraic roots of the depressed cubic. It need not coincide with a perturbative expansion along a physical background trajectory, for which both $p$ and $q$ generally evolve. We consider the downhill sector, for which \eqref{eq:cubic-branch} is satisfied:
\begin{equation}
\label{eq:cubic-app}
x^3-px+q=0 \, ,
\end{equation}
with $p>0$ and $q>0$. As shown in Section~\ref{sec:cubic}, the discriminant satisfies $\Delta=q^2/4-p^3/27\leq0$ along a physical downhill trajectory.

\paragraph{Exact roots and their identification.}
Let us define
\begin{equation}
\kappa\equiv\frac{3\sqrt{3}\,q}{2p^{3/2}} \, ,
\end{equation}
which satisfies $0\leq\kappa\leq1$ for $\Delta\leq0$. The three real roots of the depressed cubic \eqref{eq:cubic-app} can be written as the exact Cardano solutions:
\begin{equation}
\label{eq:cardano-app}
X_k=2\sqrt{\frac{p}{3}} \cos\bigg[ \frac{1}{3}\arccos(-\kappa) -\frac{2\pi k}{3} \bigg] 
\quad \text{ with } \quad
k=0,1,2 \, .
\end{equation}
We denote these roots by $X_k$ to distinguish their Cardano labels from the physical root $x_0=\sqrt{\epsilon}$ used in the main text.

For $0<\kappa<1$, the roots are ordered as
\begin{equation}
X_2<0<X_1<X_0 \, .
\end{equation}
Thus $X_1$ and $X_0$ are respectively the lower- and higher-velocity algebraic roots. The physical solution $x_0=\sqrt{\epsilon}$ coincides with one of these two roots, as determined by the background constraint. At $\kappa=1$, the discriminant vanishes and
\begin{align}
X_0 = X_1 & = \sqrt{\frac{p}{3}} \, ,
\\
X_2 & = -2\sqrt{\frac{p}{3}} \, .
\end{align}
The two positive roots therefore merge smoothly, providing the exact Cardano description of the branch-swap point.

\paragraph{Validity of the small-$q$ expansion.}
Expanding the roots of \eqref{eq:cardano-app} around $q=0$ with fixed $p>0$, we obtain
\begin{align}
X_0 
& = 
\sqrt{p} - \frac{q}{2p} + \mathcal{O} \bigg( \frac{q^2}{p^{5/2}} \bigg) 
\, ,
\\
X_1 
& = 
\frac{q}{p} + \mathcal{O} \bigg( \frac{q^3}{p^4} \bigg)
\, ,
\\
X_2 
& = 
- \sqrt{p} - \frac{q}{2p} + \mathcal{O} \bigg( \frac{q^2}{p^{5/2}} \bigg)
\, .
\end{align}
These reproduce the expansions in the main text. However, their small parameter is not simply $q$, but rather
\begin{equation}
\label{eq:smallq-validity}
\frac{q}{p^{3/2}}\ll1 \, .
\end{equation}
This distinction becomes important when $p$ also approaches zero. Indeed, at the branch-swap point,
\begin{equation}
q = \frac{2}{3\sqrt{3}}p^{3/2} \, ,
\end{equation}
so that \eqref{eq:smallq-validity} is no longer satisfied, even if $q\ll1$. Consequently the small-$q$ expansions reliably describe the two branches away from their merger, but cannot resolve the merger itself.

\paragraph{Quasi-de Sitter branch-swap regime.}
To examine this issue in terms of inflationary variables, we use the parameter $\chi=\sqrt{\epsilon_V/\epsilon}$ introduced in Section~\ref{sec:chart}. The cubic coefficients satisfy the exact relations
\begin{align}
\label{eq:pq-exact-app1}
p & = \epsilon + (3-\epsilon) \chi \, ,
\\
\label{eq:pq-exact-app2}
q & = (3-\epsilon) \chi \sqrt{\epsilon} \, .
\end{align}
For $\epsilon\ll1$, the small-$q$ expansion is valid in two asymptotic regimes, $\chi\gg\epsilon$ and $\chi\ll\epsilon$, corresponding respectively to the lower- and higher-root expansions. The familiar SR and deep USR limits lie within these respective regimes, but the correspondence does not extend to their entire domains.

We can study this region without any singular expansion by introducing
\begin{align}
c & \equiv \frac{\chi}{\epsilon} \, ,
\\
y & \equiv \frac{x}{\sqrt{\epsilon}} \, .
\end{align}
Substituting \eqref{eq:pq-exact-app1} and \eqref{eq:pq-exact-app2} into \eqref{eq:cubic-app} and dividing by $\epsilon^{3/2}$ gives the following factorized form:
\begin{equation}
\label{eq:scaled-cubic-app}
y^3 - \big[ 1 + (3-\epsilon) c \big] y + (3-\epsilon) c
=
(y-1) \big[ y^2 + y - (3-\epsilon) c \big]
=
0
\, .
\end{equation}
This factorization is {\it exact}. The physical root is $y=1$, while the other positive root is
\begin{equation}
y_+ = \frac{-1+\sqrt{1+4(3-\epsilon)c}}{2} \, .
\end{equation}
The two roots merge when $y_+=1$, or
\begin{equation}
c_{\rm swap}=\frac{2}{3-\epsilon}.
\end{equation}
Equivalently,
\begin{align}
\label{eq:swap-app1}
\chi_{\rm swap}
& =
\frac{2\epsilon}{3-\epsilon}
\, ,
\\
\label{eq:swap-app2}
\epsilon_V^{\rm swap}
& =
\frac{4\epsilon^3}{(3-\epsilon)^2}
\approx
\frac{4}{9}\epsilon^3 
\, .
\end{align}
These are precisely the condition $\Delta = 0$.

We may alternatively approximate the coefficients in \eqref{eq:pq-exact-app1} and \eqref{eq:pq-exact-app2} as $p\approx\epsilon+3\chi$ and $q\approx3\chi\sqrt{\epsilon}$. The resulting cubic still factorizes as
\begin{equation}
x^3 - (\epsilon+3\chi) x + 3\chi\sqrt{\epsilon}
=
(x-\sqrt{\epsilon}) \big( x^2 + \sqrt{\epsilon} x - 3\chi \big)
= 
0
\, .
\end{equation}
The exact appearance of $x=\sqrt{\epsilon}$ in this approximate equation follows from the consistent approximation of both coefficients. It should not be interpreted as establishing the validity of the small-$q$ series near the branch-swap point.

We conclude that the lower- and higher-velocity branches identified by the small-$q$ expansions are consistent with the exact Cardano solutions, while their merging is correctly described by the full cubic or, equivalently, by the scaled factorization \eqref{eq:scaled-cubic-app}.


\begin{thebibliography}{99}




%\cite{Guth:1980zm}
\bibitem{Guth:1980zm}
A.~H.~Guth,
%``The Inflationary Universe: A Possible Solution to the Horizon and Flatness Problems,''
Phys. Rev. D \textbf{23}, 347-356 (1981).
%doi:10.1103/PhysRevD.23.347



%\cite{Linde:1981mu}
\bibitem{Linde:1981mu}
A.~D.~Linde,
%``A New Inflationary Universe Scenario: A Possible Solution of the Horizon, Flatness, Homogeneity, Isotropy and Primordial Monopole Problems,''
Phys. Lett. B \textbf{108}, 389-393 (1982).
%doi:10.1016/0370-2693(82)91219-9



%\cite{Albrecht:1982wi}
\bibitem{Albrecht:1982wi}
A.~Albrecht and P.~J.~Steinhardt,
%``Cosmology for Grand Unified Theories with Radiatively Induced Symmetry Breaking,''
Phys. Rev. Lett. \textbf{48}, 1220-1223 (1982).
%doi:10.1103/PhysRevLett.48.1220



%\cite{Planck:2018vyg}
\bibitem{Planck:2018vyg}
N.~Aghanim \textit{et al.} [Planck],
%``Planck 2018 results. VI. Cosmological parameters,''
Astron. Astrophys. \textbf{641}, A6 (2020)
[erratum: Astron. Astrophys. \textbf{652}, C4 (2021)]
%doi:10.1051/0004-6361/201833910
[arXiv:1807.06209 [astro-ph.CO]].



%\cite{Mukhanov:1981xt}
\bibitem{Mukhanov:1981xt}
V.~F.~Mukhanov and G.~V.~Chibisov,
%``Quantum Fluctuations and a Nonsingular Universe,''
JETP Lett. \textbf{33}, 532-535 (1981).



%\cite{Guth:1982ec}
\bibitem{Guth:1982ec}
A.~H.~Guth and S.~Y.~Pi,
%``Fluctuations in the New Inflationary Universe,''
Phys. Rev. Lett. \textbf{49}, 1110-1113 (1982).
%doi:10.1103/PhysRevLett.49.1110



%\cite{Hawking:1982cz}
\bibitem{Hawking:1982cz}
S.~W.~Hawking,
%``The Development of Irregularities in a Single Bubble Inflationary Universe,''
Phys. Lett. B \textbf{115}, 295 (1982).
%doi:10.1016/0370-2693(82)90373-2



%\cite{Starobinsky:1982ee}
\bibitem{Starobinsky:1982ee}
A.~A.~Starobinsky,
%``Dynamics of Phase Transition in the New Inflationary Universe Scenario and Generation of Perturbations,''
Phys. Lett. B \textbf{117}, 175-178 (1982).
%doi:10.1016/0370-2693(82)90541-X



%\cite{Bardeen:1983qw}
\bibitem{Bardeen:1983qw}
J.~M.~Bardeen, P.~J.~Steinhardt and M.~S.~Turner,
%``Spontaneous Creation of Almost Scale - Free Density Perturbations in an Inflationary Universe,''
Phys. Rev. D \textbf{28}, 679 (1983).
%doi:10.1103/PhysRevD.28.679



%\cite{Martin:2013tda}
\bibitem{Martin:2013tda}
J.~Martin, C.~Ringeval and V.~Vennin,
%``Encyclop{\ae}dia Inflationaris: Opiparous Edition,''
Phys. Dark Univ. \textbf{5-6}, 75-235 (2014)
%doi:10.1016/j.dark.2024.101653
[arXiv:1303.3787 [astro-ph.CO]].



%\cite{Dodelson:1997hr}
\bibitem{Dodelson:1997hr}
S.~Dodelson, W.~H.~Kinney and E.~W.~Kolb,
%``Cosmic microwave background measurements can discriminate among inflation models,''
Phys. Rev. D \textbf{56}, 3207-3215 (1997)
%doi:10.1103/PhysRevD.56.3207
[arXiv:astro-ph/9702166 [astro-ph]].



%\cite{Kinney:2002qn}
\bibitem{Kinney:2002qn}
W.~H.~Kinney,
%``Inflation: Flow, fixed points and observables to arbitrary order in slow roll,''
Phys. Rev. D \textbf{66}, 083508 (2002)
%doi:10.1103/PhysRevD.66.083508
[arXiv:astro-ph/0206032 [astro-ph]].



%\cite{Tsamis:2003px}
\bibitem{Tsamis:2003px}
N.~C.~Tsamis and R.~P.~Woodard,
%``Improved estimates of cosmological perturbations,''
Phys. Rev. D \textbf{69}, 084005 (2004)
%doi:10.1103/PhysRevD.69.084005
[arXiv:astro-ph/0307463 [astro-ph]].



%\cite{Kinney:2005vj}
\bibitem{Kinney:2005vj}
W.~H.~Kinney,
%``Horizon crossing and inflation with large eta,''
Phys. Rev. D \textbf{72}, 023515 (2005)
%doi:10.1103/PhysRevD.72.023515
[arXiv:gr-qc/0503017 [gr-qc]].



%\cite{Leach:2001zf}
\bibitem{Leach:2001zf}
S.~M.~Leach, M.~Sasaki, D.~Wands and A.~R.~Liddle,
%``Enhancement of superhorizon scale inflationary curvature perturbations,''
Phys. Rev. D \textbf{64}, 023512 (2001)
%doi:10.1103/PhysRevD.64.023512
[arXiv:astro-ph/0101406 [astro-ph]].



%\cite{Byrnes:2018txb}
\bibitem{Byrnes:2018txb}
C.~T.~Byrnes, P.~S.~Cole and S.~P.~Patil,
%``Steepest growth of the power spectrum and primordial black holes,''
JCAP \textbf{06}, 028 (2019)
%doi:10.1088/1475-7516/2019/06/028
[arXiv:1811.11158 [astro-ph.CO]].



%\cite{Motohashi:2014ppa}
\bibitem{Motohashi:2014ppa}
H.~Motohashi, A.~A.~Starobinsky and J.~Yokoyama,
%``Inflation with a constant rate of roll,''
JCAP \textbf{09}, 018 (2015)
%doi:10.1088/1475-7516/2015/09/018
[arXiv:1411.5021 [astro-ph.CO]].



%\cite{Pattison:2018bct}
\bibitem{Pattison:2018bct}
C.~Pattison, V.~Vennin, H.~Assadullahi and D.~Wands,
%``The attractive behaviour of ultra-slow-roll inflation,''
JCAP \textbf{08}, 048 (2018)
%doi:10.1088/1475-7516/2018/08/048
[arXiv:1806.09553 [astro-ph.CO]].



%\cite{Prokopec:2025uvz}
\bibitem{Prokopec:2025uvz}
T.~Prokopec and G.~Rigopoulos,
%``Inflaton perturbations through an ultra-slow-roll transition and Hamilton-Jacobi attractors,''
JCAP \textbf{04}, 028 (2026)
%doi:10.1088/1475-7516/2026/04/028
[arXiv:2507.04114 [gr-qc]].



%\cite{Kusmartsev:1994pz}
\bibitem{Kusmartsev:1994pz}
F.~V.~Kusmartsev, E.~W.~Mielke, Y.~N.~Obukhov and F.~E.~Schunck,
%``Classification of inflationary Einstein scalar field models via catastrophe theory,''
Phys. Rev. D \textbf{51}, 924-927 (1995)
%doi:10.1103/PhysRevD.51.924
[arXiv:gr-qc/9412046 [gr-qc]].



%\cite{Planck:2015sxf}
\bibitem{Planck:2015sxf}
P.~A.~R.~Ade \textit{et al.} [Planck],
%``Planck 2015 results. XX. Constraints on inflation,''
Astron. Astrophys. \textbf{594}, A20 (2016)
%doi:10.1051/0004-6361/201525898
[arXiv:1502.02114 [astro-ph.CO]].



%\cite{Romano:2015vxz}
\bibitem{Romano:2015vxz}
A.~E.~Romano, S.~Mooij and M.~Sasaki,
%``Adiabaticity and gravity theory independent conservation laws for cosmological perturbations,''
Phys. Lett. B \textbf{755}, 464-468 (2016)
%doi:10.1016/j.physletb.2016.02.054
[arXiv:1512.05757 [gr-qc]].



%\cite{Vallejo-Pena:2019lfo}
\bibitem{Vallejo-Pena:2019lfo}
S.~A.~Vallejo-Pe{\~n}a and A.~E.~Romano,
%``Are primordial black holes produced by entropy perturbations in single field inflationary models?,''
JCAP \textbf{11}, 015 (2019)
%doi:10.1088/1475-7516/2019/11/015
[arXiv:1904.07503 [astro-ph.CO]].



%\cite{Starobinsky:1992ts}
\bibitem{Starobinsky:1992ts}
A.~A.~Starobinsky,
%``Spectrum of adiabatic perturbations in the universe when there are singularities in the inflation potential,''
JETP Lett. \textbf{55}, 489-494 (1992).



%\cite{Wands:1998yp}
\bibitem{Wands:1998yp}
D.~Wands,
%``Duality invariance of cosmological perturbation spectra,''
Phys. Rev. D \textbf{60}, 023507 (1999)
%doi:10.1103/PhysRevD.60.023507
[arXiv:gr-qc/9809062 [gr-qc]].




\end{thebibliography}
\end{document}